\documentclass{article}
\usepackage{spconf,amsmath,graphicx,url}
\usepackage{adjustbox, booktabs}
\ninept
\title{Fine-tuning Strategies for Faster Inference using Speech Self-Supervised Models:  A Comparative Study}
\name{Salah Zaiem$^{\star,  \mp}$ \qquad Robin Algayres$^{\diamond}$ \qquad Titouan Parcollet$^{\dagger}$ \qquad Slim Essid$^{\star}$ \qquad Mirco Ravanelli$^{\mp}$ }
\address{$\star$ Telecom Paris, Palaiseau, France, $\diamond$ ENS, INRIA, INSERM, UPEC, PSL Research University, France \\
 $\dagger$ Samsung AI Center, Cambridge, United-Kingdom, \\
 $\mp$ Mila-Quebec AI Institute, Université de Montréal, Concordia University, Canada}
\begin{document}
%\ninept
%
\maketitle
\begin{abstract}
Self-supervised learning (SSL) has allowed substantial progress in Automatic Speech Recognition (ASR) performance in low-resource settings. In this context, it has been demonstrated that larger self-supervised feature extractors are crucial for achieving lower downstream ASR error rates. Thus, better performance might be sanctioned with longer inferences. This article explores different approaches that may be deployed during the fine-tuning to reduce the computations needed in the SSL \textit{encoder}, leading to faster inferences. We adapt a number of existing techniques to common ASR settings and benchmark them, displaying performance drops and gains in inference times. Interestingly, we found that given enough downstream data, a simple downsampling of the input sequences outperforms the other methods with both low performance drops and high computational savings, reducing computations by $61.3\%$ with an WER increase of only $0.81$. Finally, we analyze the robustness of the comparison to changes in dataset conditions, revealing sensitivity to dataset size.

\end{abstract}
\begin{keywords}
Speech recognition, self-supervised learning.
\end{keywords}
\section{Introduction}
\label{sec:intro}
Self-supervised learning (SSL) has emerged as the main approach for leveraging unlabelled data to achieve significant performance improvements in a wide range of downstream tasks. This is particularly beneficial as it reduces the need for expensive and imprecise manual annotation. Various approaches have been introduced in the literature, including predictive coding, \cite{baevski2020wav2vec}, multi-task learning \cite{ravanelli2020multitask, zaiem2022pretext}, auto-encoding techniques \cite{algayres2020evaluating} or contrastive learning \cite{zaiem2022automatic}.

However, recent trends in SSL for speech have shown that the improvements in terms of performance are often driven by larger architectures, leading to potentially long inference times \cite{Chen2021}. For instance, Sanyuan et al. \cite{Chen2021}, have shown that switching from WavLM Large to WavLM Base halved the observed word error rate (WER) on a held-out English ASR task. Preserving reasonable inference times while increasing the capacity of the model is of critical interest to maximize the impact of this new technology on real-life products.

As a matter of fact, several approaches have been proposed to shorten inference times using SSL models. Some attempted to distill state-of-the-art models by using shallower or thinner networks \cite{Chang2021, Wang} or through downsampling the inputs \cite{Chen2022, lee2022fithubert}. However, while the downstream performance of distilled student models is comparable to larger teacher models on most speech classification tasks, a large gap is still witnessed for more complex tasks such as ASR \cite{Feng2022}. Also, low-bit quantization during pretraining has recently emerged as a successful approach for faster inference times \cite{Yeh}. Compared to our proposed methods, these two approaches bear the advantage of leading to generalist models usable for multiple downstream tasks. However, they have two major downsides. First, they necessitate access to the very large pretraining dataset, which may or may not be publicly and commercially available, such as for recent and large-scale state-of-the-art speech recognition models \cite{whisper}. Second, even if the pretraining set is available, applying quantization or distillation to these large models remains a particularly challenging and costly task due to the original dataset and model sizes. For instance, academic attempts for distilling SSL models have been solely applied to Base models (\textit{i.e} less than 100M parameters), and are generally restricted to a thousand hours of speech pretraining data, compared to $94k$ hours for 317M parameters WavLM Large.

In contrast, by focusing on a given downstream task of interest, our approach allows for reducing the inference time using off-the-shelf large self-supervised models, directly training only on the small considered downstream dataset. This is achieved by benefiting from the fine-tuning phase to reduce the inference time. More precisely, this article considers three families of techniques originating from the recent SSL literature: i) layer dropping or replacement, ii) early-exiting, and iii) input sequence downsampling. The fine-tuning phase allows the models to adapt to a shrunk architecture, or shorter inputs while maximizing the downstream performance. Thus, our contributions are twofold: i)We compare a number of recent shrinking approaches (Section \ref{sec:methods}) used while fine-tuning large SSL models on an ASR task providing inference speed metrics, and ii) we analyze the robustness of these methods to changes in the dataset and in the annotated data quantity (Section \ref{sec:robustness}). The SpeechBrain code base is released for replication and further advancements \cite{speechbrain}. \footnote{\url{https://github.com/salah-zaiem/speeding_inferences}}
%\end{enumerate}

\section{Setting and Methods}
\label{sec:methods}
This section outlines the global setting for comparing the considered techniques before providing a detailed description of the approaches and their motivations.

\subsection{Benchmarking setting}\label{sec:setting}
The study is conducted under the  two strong yet realistic following conditions. First, we suppose that we do not have access to the pretraining dataset that would enable, for instance, extensive distillation or quantization approaches.  Second, we limit ourselves to using only the annotated training data of the target dataset during fine-tuning, eliminating transfer-learning based approaches. The second condition is relevant when the target domain is rare and specific enough not to take advantage of transfer learning from classic large annotated datasets. This is generally true in two popular cases where using self-supervised models is privileged: low-resourced languages  and speech datasets with specific acoustic conditions \cite{zuluaga}.

We use the released pre-trained and non fine-tuned WavLM Large \cite{Chen2021} as the SSL model, as it tops speech self-supervised learning benchmarks and exhibits resilience to noisy conditions \cite{Feng2022}. In all the experiments of this section, we use the \textit{train-clean-100} split of LibriSpeech \cite{libri} as our training set, the \textit{dev-clean} split for validation and finally the \textit{test-clean} split for testing. Following common practices \cite{hubert}, we freeze the convolutional front-end and only fine-tune the transformers part of  WavLM Large consisting of 24 transformers layers. The self-supervised \textit{encoder} outputs a frame vector of dimension $1,024$ for every segment of 320 speech samples which corresponds in the case of a 16-kHz sampling rate to 20 ms of audio signal. Two fully connected layers with a hidden size of $1,024$ map each frame vector to the probabilities of the considered characters. Connectionist Temporal Classification (CTC) \cite{ctc} loss is used for training. 

\begin{table*}[]
\centering
 \vspace{-14mm}
 \scalebox{0.64}{\begin{tabular}{lllllllll}
\toprule
\textbf{Technique}                         & \textbf{}                      &\multicolumn{1}{l}{\textbf{WER  $\downarrow$}} & \multicolumn{1}{l}{\textbf{GPU (s)}} & \multicolumn{1}{l}{\textbf{CPU (s)}} & \multicolumn{1}{l}{\textbf{WER-LM  $\downarrow$}} & \multicolumn{1}{l}{\textbf{GPU-LM (s)}} & \multicolumn{1}{l}{\textbf{CPU-LM (s)}} & \textbf{MACs (G)}           \\\midrule
\textbf{Baseline}            & \textbf{Full Model}              & 4.09                             & 134                                  & 1121                                 & 3.31                                & 152                                     & 1128                                    & 386.53    \\  \midrule \midrule
\textbf{Layer Drop}                        & \textbf{Drop Prob}             & \multicolumn{1}{l}{}             & \multicolumn{1}{l}{}                 & \multicolumn{1}{l}{}                 & \multicolumn{1}{l}{}                & \multicolumn{1}{l}{}                    & \multicolumn{1}{l}{}                    &                             \\\midrule
                                           & 0.5                            & 11.28                            & 96                                   & 721                                  & 5.89                                & 156                                     & 776                                     & 244.19                   \\
                                           & 0.4                            & 8.32                             & 102                                  & 816                                  & 4.58                                & 145                                     & 844                                     & 272.28                    \\
                                           & 0.3                            & 6.56                             & 109                                  & 888                                  & 3.84                                & 157                                     & 913                                     & 300.98                     \\
                                           & 0.25                           & 5.91                             & 113                                  & 932                                  & 3.72                                & 148                                     & 950                                     & 314.24                     \\ \midrule
\textbf{Layer Removal}                     & \textbf{Num. Kept Layers}           &                                  &                                      &                                      &                                     &                                         &                                         &                             \\\midrule
\multicolumn{1}{r}{}                       & \multicolumn{1}{l}{12}         & 14.39                            & 93                                   & 726                                  & 8.64                                & 127                                     & 739                                     & 236.64                     \\
\multicolumn{1}{r}{}                       & \multicolumn{1}{l}{16}         & 8.16                             & 109                                  & 852                                  & 5.53                                & 131                                     & 861                                     & 286.60                   \\
                                           & \multicolumn{1}{l}{20}         & 5.14                             & 117                                  & 988                                  & 3.62                                & 142                                     & 989                                     & 336.57               
               \\\midrule \midrule
\textbf{Early Exit : Entropy Threshold}                     & \textbf{Mean Exit Layer}           &                                  &                                      &                                      &                                     &                                         &                                         &                             \\\midrule
0.06    & 13.80 & 12.08 & 96                       & 757                       &  9.25   & 122                        & 765         & 252.36              \\ 
0.03     & 17.61       & 7.67                & 116                      & 974                      & 6.55   & 137                        & 976     & 326.28                    \\ 
0.025    & 20.52         & 6.66              & 128                      & 1127                     & 5.87   & 149                        & 1132    & 364.92                   \\ 
0.01      & 23.98     & 6.20                 & 142                      & 1275                     & 5.49   & 165                        & 1280         & 386.53              \\\midrule 

\textbf{Early Exit :  Layer Sim. Threshold}     &   \textbf{Mean Exit Layer}     &                            &   &    &          &        &      &     \\ \midrule
0.92    & 15.97    &    10.23                & 99                       & 812                       & 8.17   & 123                        & 819 & 274.11                       \\ 
0.95    &  17.18   &  8.78                   & 104                      & 850                      & 7.35   & 126                        & 864   & 291.68                      \\ 
0.965    & 21.44    & 6.79                   & 120                      & 1070                     & 5.93   & 131                        & 1073    & 358.85                    \\ 
0.98     & 24.00       & 6.20                & 1280                     & 1153                     & 5.49   & 149                        & 1153   & 386.51                     \\ \midrule 

\textbf{Two Steps EE :  Layer Sim. Threshold}     &   \textbf{Mean Exit Layer}     &                            &   &    &          &        &      &     \\ \midrule
0.96                          & 14.52                & \multicolumn{1}{l}{21.95}               & 102                              & 866                              & \multicolumn{1}{l}{8.75}            & 180                                & 938                                & 285.68              \\
0.97                          & 21.46                & \multicolumn{1}{l}{6.17}                & 126                              & 1138                             & \multicolumn{1}{l}{4.34}            & 152                                & 1167                               & 382.00              \\
0.98                          & 23.0                 & \multicolumn{1}{l}{4.54}                & 130                              & 1175                             & \multicolumn{1}{l}{3.87}            & 151                                & 1196                               & 386.54              \\
  \midrule \midrule

\multicolumn{1}{l}{\textbf{Downsampling Technique}} & \multicolumn{1}{l}{\textbf{Downsampling Factor}}  \\ \midrule
 \multicolumn{1}{l}{\textbf{Convolutional Downsampling}}                                    &2                                      & 4.61                             & 84                               & 582                              & 3.48                                & 98                                 & 600                                & 192.97                            \\

 & \textbf{3} & \textbf{5.47} & \textbf{69} & \textbf{414} & \textbf{4.12} & \textbf{91} & \textbf{436}& \textbf{134.86} \\
 &4&21.88&67&335&14.60&106&340&96.11\\ \midrule
 
\multicolumn{1}{l}{\textbf{Averaging Downsampling}}  
                                    & 2                                      & 4.93                             & 80                               & 570                              & 3.66                                & 98                                 & 578                                & 192.97                            \\
                                      & 3                                      & 6.01                             & 64                               & 406                              & 4.27                                & 90                                 & 422                                & 134.86                         \\
                                      & 4                                      & 26.84                            & 60                               & 326                              & 18.02                               & 115                                & 385                                & 96.11                             \\\midrule
\multicolumn{1}{l}{\textbf{Signal Downsampling}}  
                                      & 2                                      & 4.85                             & 86                               & 569                              & 3.58                                & 97                                 & 575                                & 192.97                            \\
                                      & 3                                      & 5.83                             & 72                               & 427                              & 4.08                                & 89                                 & 458                                & 134.86                         \\
                                      & 4                                      & 16.08                            & 63                               & 330                              & 11.10                                & 97                                 & 369                                & 96.11          \\ \midrule \midrule
                \multicolumn{1}{l}{\textbf{DistilHuBERT}} & \multicolumn{1}{l}{Linear Decoder} & 30.74                            & 56                                   & 240                                  & 16.20                                & 130                                     & 311                                     & 101.74                     \\
                                           & BiLSTM Decoder                        & 16.30                            & 95                                   & 545                                  & 10.57                               & 128                                     & 613                                     & 161.06   \\ \bottomrule 

\end{tabular}}
\caption{Word error rates and inference metrics on LibriSpeech \textit{test-clean} split for the considered approaches and various parameters per method. All models are finetuned on LibriSpeech \textit{train-clean-100}. ``GPU" and ``CPU" indicate the inference times in seconds on GPU and CPU. ``-LM" suffixes indicate that the decoding uses a language model. ``Drop Prob" is the probability of randomly dropping layers during inference.  Early-exiting experiments come with an exiting threshold and a resulting mean exit layer computed over the test set.}
\label{tab:layers}

\end{table*}

\iffalse
\multicolumn{1}{l}{\textbf{FTMelWavLM}}      & \multicolumn{1}{l}{}           & 12.84                            & 120                                  & 945                                  & 9.73                                & 142                                     & 928                                     & 314.15 \\ \midrule \midrule
\fi
During inference, the decoding of the probabilities of the characters is completed in two ways: with or without using a Language Model (LM). In the experiments labeled as \textit{Without LM}, greedy decoding is applied, outputting the character with the maximal probability at each step before applying CTC-based reformatting to get the predicted words. In the \textit{With LM} experiments, we use the Librispeech official 4-gram language model, trained using the KenLM library \cite{heafield-2011-kenlm}, and decode the sentence using shallow fusion \cite{shallow} considering the language modelling score of a beam of the most acoustically probable sequences. An n-gram model is chosen over more complex language modelling approaches to reduce word error rates while keeping low inference times. This is done using the PyCTCDecode \footnote{\url{https://github.com/kensho-technologies/pyctcdecode}} library with default parameters. To understand the impact of this aspect on the results tables, it is crucial to note that this decoding is performed on CPU and that the decoding time is prolonged when the model is uncertain about its predictions. Indeed, PyCTCDecode proceeds to prune elements of the beam that are scored too low by the language model compared to the maximal beam score. It leads in this work to a penalty for models with high error rates before the LM addition, as they systematically had longer decoding times. With the stage of the comparison set, we will proceed to the descriptions of the selected candidates

\subsection{Layer dropping and replacement}
\label{sec:layerdrop}
With multiple studies on layer-wise probing of self-supervised models showing that phonetic content is divided among the layers of the transformers \cite{osti_10303839}, removing layers has emerged as a possibility for faster inferences. Experiments led mainly on text language models have shown that dropping higher level layers is preferable to avoid heavy performance drops \cite{Sajjad_2023}. In a first experiment, we will study the effect of fine-tuning the SSL model after having removed a number of layers. In a second one, given the fact that WavLM has been trained with \textit{layerdrop} \cite{layerdrop}, \textit{i.e} random layer omission during training, we fine-tune it with  layerdrop probability equal to $q=0.5$ and study the effect of keeping various layerdrop rates during test. 
\iffalse
Finally, inspired by MelHubert \cite{melhubert}, we test replacing the convolutional front-end of WavLM with Mel spectrograms during the fine-tuning, we call this experiment ``FTMelWavLM". 
\fi

\subsection{Early-exiting}
\label{sec:early}

 Similarly, early-exiting is a relevant approach to reduce computations during inference \cite{overthinking, Yoon}. It consists in allowing the model to use an early layer representation and feed it directly to the decoder, saving the computation of further layers. During fine-tuning, starting from the twelfth layer, a specific downstream decoder is learned on top of every layer. During inference, a heuristic metric computed after each layer indicates whether the model should output a decoded sequence using the current layer or go further. Given well calibrated heuristics, early-exiting should reduce the mean exit layer and, thus, the mean inference time. Furthermore, studies on what has been called ``overthinking" \cite{overthinking} have shown that SSL models could benefit from early exiting both inference time and performance. Inspired by previous works, two heuristics are tested: the entropy of the decoder outputs, and a measure of similarity between the representations of consecutive layers. 

 As described in Section \ref{sec:setting}, each downstream decoder consists of two linear layers outputting logit probabilities after each time frame of 20 ms. Each layer $i$ of the transformer outputs  a representation $R_i$ of size  $[N,A]$ with $N$ the number of signal time frames and $A$ the hidden dimension of WavLM Large ($A = 1,024$). Operating on this representation through the decoder $D_i$, the vectors of logits probabilities $L_i$ are of size $[N,P]$ with $P$ the number of different characters in the dataset (in our case $P=31$). The entropy $E_i$ of the output of layer $i$ is defined as: 
 \begin{equation}
\label{entropy}
E_i = -\frac{1}{NP}\sum_{j=1}^{N}  \sum_{k=1}^{P} L_{i,j,k} \log(L_{i,j,k}).
\end{equation}

with $L_{i,j,k}$ being the probability of the character $k$ at time frame $j$ predicted by decoder $D_i$. During fine-tuning, to learn the weights of every decoders weights, we pass through all the  layers of the model and sum the CTC losses over the outputs of all decoders. During inference, after each layer $i$ starting from the twelfth, $R_i$ is decoded and $E_i$ is computed. If $E_i$ is lower than a fixed entropy threshold, we do not go further in the SSL transformer, and decode the logit probabilities into words. Hence, reducing the entropy threshold increases the confidence required for exiting, leading systematically in this work to later exits and thus, higher inference times.  
 The second exiting heuristics is layer representation similarity. For each layer $i \geq 12$, we compute the cosine similarity between $R_i$ and $R_{i-1}$. Similarly to the first approach, if the similarity is higher than a fixed threshold at layer $i$, the model exits to $D_i$ and decodes the logits into a word sequence. This second approach is slightly faster as it does not involve computing the decoding into logits at each layer. 

\subsection{Sequence Downsampling}
\label{sec:downsampling}
Inspired by works on distilling speech models with smaller sampling rates \cite{Wang, lee2022fithubert}, and given the quadratic memory bottleneck of transformers architectures as a function of input lengths, we assess the capacity of the SSL model, trained on 16-kHz audio inputs, to adapt to lower sampling rates. Given a speech file $x$ consisting in $T$ speech samples $x=(x_i)_{i \in [1,T]}$ and a downsampling factor $k$, a function $f$, learned or unlearned depending on the chosen method, downsamples $x$ to $x' = f(x) = ({x'_i})_{i \in [1, \lfloor T/k \rfloor]}$, a sequence of size ${\lfloor T/k \rfloor}$. The downsampled sequence $x'$ is then fed to the SSL feature extractor instead of $x$. Three methods for downsampling the input sequences are evaluated. The first one is signal decimation (\textit{i.e.} classic signal downsampling). Second, we test a learned downsampling strategy, through a one-dimensional (1D) convolution layer of kernel size 160 and a stride equal to the downsampling factor, ran on the input waveform. Finally, we test an averaging 1D downsampling with a fixed window of size 16 (\textit{i.e.} a constant convolution).

Each one of the techniques is evaluated with 3 downsampling factors: 2, 3 and 4. For instance, this corresponds for the first approach to downsampling the signals from 16 kHz to, respectively, 8000, 5333, and 4000 Hz. As explained in Section \ref{sec:setting}, the SSL model outputs a character every 320 audio samples, which corresponds to 20 ms of audio with a 16-kHz sampling rate. With lower sampling rates, the number of output characters may become lower than the lengths of the corresponding textual sequences. This is why, when dealing with downsampling factors 3 and 4, the size of the decoder output layer is doubled. It is  then reshaped to fit the number of considered characters, before being fed to the classifier softmax function. This allows every frame of audio to output two characters instead of one.
\vspace{-2mm}

\section{Results and Robustness Study}
\label{sec:robustness}

 Table \ref{tab:layers} shows the results obtained with the different techniques. Reported GPU inference times are for a Nvidia Tesla V100 SXM2 32 Go GPU, while CPU inferences are using  one Intel Cascade Lake 6248 processor with a 27.5 MB cache and 2.50 GHz clock speed. The inference times and MACs are those for running inference on the whole $5.4$ hours of the \textit{test-clean} split. 
 
 \noindent \textbf{Layer removals.} The results of the layer removal and dropping approaches are displayed in the upper part of Table \ref{tab:layers}. Surprisingly, for a given proportion of layers dropped, keeping the layerdrop performs better than training with the reduced number of layers. For instance, randomly dropping $50\%$ of the layers during the test leads to $11.28$ of WER, compared to $14.39$ when dropping the last 12 layers during fine-tuning(\textit{i.e.} again $50\%$ of the layers). It suggests that while training systematically with the same layers adapts the models directly to inference conditions, removing the information contained in the last layers of the model harms too severely the performance.
 
\begin{figure*}[ht!]
  \centering
  \includegraphics[width=0.9\linewidth, scale=0.2]{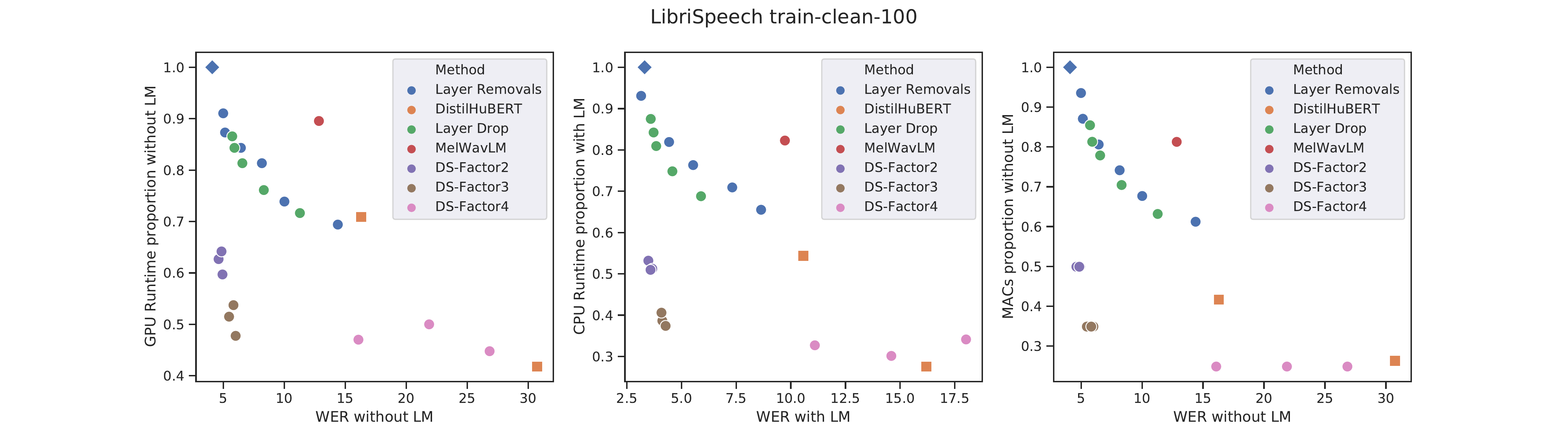}
   \caption{WER and inference metrics with or without language modelling for the presented techniques fine-tuned on LibriSpeech-100h. The best techniques, characterized by both low Word Error Rates (WERs) and inference times, are Factor2 and Factor3 downsamplings, located in the bottom left of the figures. The full model is indicated by a blue diamond, while DistilHubert baselines are represented by orange squares. Inference time measurements are shown as a proportion of the measure done with the full model.}
  \label{fig:libri}

\end{figure*}

\begin{figure*}[ht!]
  \centering
    
  \includegraphics[width=0.9\linewidth, scale=0.2]{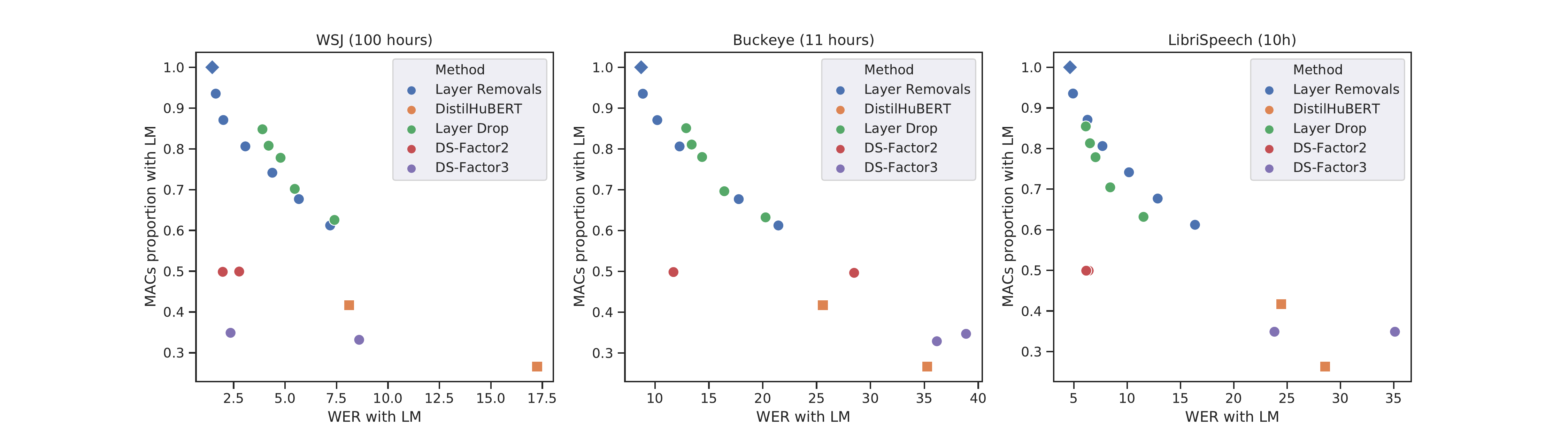}
  \caption{WER with LM decoding and MACs for the considered methods on WSJ, Buckeye and LibriSpeech-10h sets. While WSJ exhibits results similar to LibriSpeech,reducing the quantity of fine-tuning data causes significant performance drops for the downsampling methods.}
\label{triptique}
\vspace{-1em}
\end{figure*}

  \noindent\textbf{Early-exiting.} The middle part of Table \ref{tab:layers} shows the obtained results using early-exiting with different values of entropy and similarity thresholds. Increasing the entropy threshold (or decreasing the similarity one) leads naturally to earlier exits (lower ``Mean Exit Layer" values) and reduced inference times but higher WERs. Results show that using our two considered heuristics to control exiting does not prevent significant  performance drops in the case of earlier exits. We can also observe that even when using the whole network (\textit{i.e} low entropy cases), this technique leads to lower performances compared to the full model trained without early exiting. We suggest the following explanation: since early exits encourage the model to push the phonetic content required for decoding towards early layers, it undermines the ability of the model to learn hierarchical features, ultimately resulting in poorer performance even when exiting later in the model. To verify this explanation, we propose a two-step approach where the SSL model weights are first fine-tuned without early-exits, before freezing them when learning the early-exit decoders. As shown in Table \ref{tab:layers}, this leads to good performances in case of late-exiting, but with the cost of steeper drops when exiting earlier. This suggests that successful early-exiting should decouple hierarchical feature extraction and decoding preparation. In both cases, early exiting lags behind layer-removal techniques in terms of ratio inference gains / performance drop.

 \noindent \textbf{Downsampling.} Results of the downsampling experiments are shown in the last part of Table \ref{tab:layers}. Downsampling by factors 2 and 3 lead to high gains in inference times without significant drops in performance. For instance, compared to running the full model, signal downsampling with a factor 3 using a language model for decoding, leads to $61.34\%$ relative CPU inference time reduction, with an absolute increase of only $0.81$ in WER. Downsampling with factor 4, while naturally leading to further gains in inference times, results in intolerable performance costs. The three considered downsampling strategies are very similar in terms of error rates and computational savings, with a slight advantage for convolutional downsampling when sequences lengths are reduced with factors 2 and 3. For comparison with baselines, we add two experiments using DistilHuBERT \cite{Chang2021}. When using the simple linear decoder used in this benchmark, DistilHuBERT shows performances largely below the ones in the original paper. For a fair comparison, we produced an experiment with a BiLSTM decoder. While improving largely the performance, this comes at a high cost in terms of inference times. 

Figure \ref{fig:libri} presents a visual comparison between all the presented methods. Clearly, factor 2 and 3 downsampling techniques are the best performing methods with low WER, jointly with low GPU and CPU inference times. While being the fastest, higher downsampling factors and DistilHuBERT suffer from high error rates.
\vspace{-1em}
\subsection{Robustness to changes in the downstream dataset} 
\vspace{-0.64em}

Finally, we test the  robustness of these conclusions to changes in the characteristics of the downstream dataset. Three datasets are considered. We tested the same methods with a 100-hour subset of the Wall Street Journal (WSJ) dataset \cite{paul-baker-1992-design} \footnote{We combined WSJ0 and WSJ1, 70-hour long each, and removed all utterances that contain non-letter symbols in their transcriptions. Then, we extracted a 100-hour random subset of the remaining sentences} . We also test the robustness of the approach to dataset size variation by reducing the fine-tuning dataset to LibriSpeech-10h train set in first experiment and training on a small spontaneous English dataset, the Buckeye corpus \cite{buckeye} containing 11 hours of data, in a final one.

Figure \ref{triptique} shows the performance obtained with the presented methods on the three datasets. While WSJ shows very similar patterns to the first set of experiments, we can see in the case of less fine-tuning data that the downsampling method performance drops significantly. Downsampling with factor 3, while suffering a relative WER augmentation of only $33.7\%$ for Librispeech-100h and $39.1\%$ for WSJ, witnesses a drop of $384.9\%$ with Buckeye and $571.7\%$ with LibriSpeech-10h compared to the full model performance. In contrast, the other methods seem more resilient to reduced data quantity. Despite this, downsampling the sequences by a factor 2 using a learned convolution remains a good option for highly reduced inference times without unacceptable performance drop.
\vspace{-1em}
\section{Conclusion}
\vspace{-1em}
In this work, we explored different methods to reduce speech recognition inference times using large self-supervised models through fine-tuning. The comparison of these methods indicates that sequence downsampling is the best performing option allowing substantial computation gain with low performance drops. Experiments led on other downstream datasets show that the size of the downstream dataset is critical to avoid  high error rates.
\section{Acknowledgements}
This work has benefited from funding from l'Agence de l'Innovation de Défense, and was performed using HPC resources from GENCI-IDRIS (Grant 2021-AD011012801R1).

\bibliographystyle{IEEEbib}
\bibliography{mybib}

\end{document}